\documentclass[10pt,fullpage]{article}

\usepackage{fullpage}
\usepackage{booktabs} 
\usepackage{array}
\usepackage{listings}
\usepackage{algorithm,algpseudocode}
\usepackage{float}
\usepackage{subcaption}
\usepackage{color,soul}
\usepackage{multirow,makecell}
\usepackage{acronym} 
\usepackage{hhline}
\usepackage{enumitem}
\usepackage{comment}
\usepackage{amsmath}
\usepackage{url}
\usepackage{hyperref}
\usepackage{times}
\usepackage{multicol}
\usepackage{adjustbox}

\usepackage{lipsum} 
\usepackage{xargs} 
\usepackage{xcolor}
\usepackage[colorinlistoftodos,prependcaption,textsize=tiny]{todonotes}

\newcommandx{\unsure}[2][1=]{\todo[linecolor=red,backgroundcolor=red!25,bordercolor=red,#1]{#2}}
\newcommandx{\change}[2][1=]{\todo[linecolor=blue,backgroundcolor=blue!25,bordercolor=blue,#1]{#2}}

\newcommandx{\info}[2][1=]{\todo[linecolor=OliveGreen,backgroundcolor=width=13cm,height=4cmOliveGreen!25,bordercolor=OliveGreen,#1]{#2}}
\newcommandx{\improvement}[2][1=]{\todo[linecolor=Plum,backgroundcolor=Plum!25,bordercolor=Plum,#1]{#2}}
\newcommandx{\thiswillnotshow}[2][1=]{\todo[disable,#1]{#2}}

\algnewcommand{\LineComment}[1]{\State \(\triangleright\) #1}

\makeatletter
 {\vskip 5pt\begin{list}{}{%
  \setlength{\topsep}{0pt}\setlength{\partopsep}{0pt}\setlength{\parskip}{0pt}%
  \setlength{\parsep}{0pt}\setlength{\labelwidth}{0pt}%
  \setlength{\rightmargin}{0pt}\setlength{\leftmargin}{0pt}%
  \setlength{\labelsep}{0pt}%
  \obeylines\@vobeyspaces\normalfont\ttfamily%
  \item[]}}
 {\end{list}\vskip5pt\noindent}
\makeatother

\begin{document}

\date{November 2021}

\title{Simpli-Squared: A Very Simple Yet Unexpectedly Powerful Join Ordering Algorithm Without Cardinality Estimates}

\author{
Asoke Datta, Yesdaulet Izenov, Brian Tsan, Florin Rusu\\
\{adatta2, yizenov, btsan, frusu\}@ucmerced.edu\\
University of California Merced
}

\maketitle

\begin{abstract}

The Join Order Benchmark (JOB) has become the de facto standard to assess the performance of relational database query optimizers due to its complexity and completeness. In order to compute the optimal execution plan -- join order -- existing solutions employ extensive data synopses and correlations -- functional dependencies -- between table attributes. These structures incur significant overhead to design, build, and maintain. In this paper, we present \textit{Simpli}city \textit{Simpli}fied (\textit{Simpli-Squared}), a very simple join ordering algorithm that achieves unexpectedly good results. Simpli-Squared computes the join order without using any statistics or cardinality estimates. It takes as input only the referential integrity constraints declared at schema definition and the number of tuples (size) in the base tables. The join order of a given query is computed by splitting the join graph along the many-to-many joins and sorting the tables based on their size. The tables involved in one-to-many joins are greedily included based on size and the query join graph. The resulting plan can be efficiently generated by a lightweight query rewriting procedure. Experiments on the JOB benchmark in PostgreSQL show that Simpli-Squared achieves runtimes having an increase of only up to 16\% -- and sometimes even a reduction -- compared to four state-of-the-art solutions that are considerably more intricate. Based on these results, we question whether JOB adequately tests query optimizers or if accurate cardinality estimation is such a fundamental requirement for performing well on the JOB benchmark.

\end{abstract}

\section{INTRODUCTION}\label{sec:intro}

Since its introduction in~\cite{Leis:QOREALLY:pvldb-2015} -- and especially after its refinement~\cite{Leis:JOB:vldb-2018} -- the JOB benchmark has become the de facto standard to assess the performance of relational database query optimizers---including cardinality estimation~\cite{Leis:index-join-sample:cidr-2017,Kiefer:KDE:pvldb-2017,Kipf:LCECJDL:cidr-2019,Hilprecht:DeepDB:pvldb-2020,Yang:NeuroCard:pvldb-2021} and join order optimization~\cite{Cai:PCETUB:sigmod-2019,Hertzschuch:simplicity:cidr-2021,Izenov:compass:sigmod-2021}. The JOB benchmark is comprehensive. It consists of 113 queries divided into 33 families varying the difficulty of the selection predicates on the 21 base tables connected by an intricate schema. The queries are complex. They include multiple instances of a table and cycles. Their number of join predicates ranges from 4 to 28. The benchmark data are realistic, including skewed attributes and cross-table correlations. Thus, there is no surprise that traditional query optimizers making unrealistic assumptions about data have a hard time performing well on JOB~\cite{Leis:JOB:vldb-2018,Izenov:compass:sigmod-2021}.

\smallskip
\noindent
Consequently, a series of novel methods targeted at JOB have been proposed. They can be divided into two categories. The methods in the first category~\cite{Cai:PCETUB:sigmod-2019,Hertzschuch:simplicity:cidr-2021} enhance cardinality estimation with functional dependency information declared in the JOB schema  to steer the join ordering algorithm toward better plans. The second category includes learning-based methods~\cite{Marcus:NEO:pvldb-2019,Negi:FlowLoss:pvldb-2021} that predict the optimal join order based on a model pre-trained over a large -- and related -- query workload. While all these methods are shown to outperform standard query optimizers, only \textit{Simplicity}~\cite{Hertzschuch:simplicity:cidr-2021} is fully-integrated in PostgreSQL~\cite{postgres} without incurring any preprocessing overhead.

\smallskip
\noindent
The motivation for our work lies in the experiments performed with the MapD in-memory database~\cite{mapd} -- currently OmniSci -- for the COMPASS sketch-based optimizer~\cite{Shin:ESC:arxiv-2019,Izenov:compass:sigmod-2021}. The MapD query optimizer is simplistic. It computes the join order by sorting the tables in decreasing order of their size. No statistics or any other information is used. As expected, the analytical cost -- the sum of intermediate cardinalities -- of the resulting plan is orders of magnitude worse than the optimal cost. However, MapD performs unexpectedly well for the JOB queries with less than 15 join predicates. While this can be attributed to the highly-optimized MapD execution engine that uses extensive multithreading and vectorization, there are 14 JOB queries for which the MapD plans executed in PostgreSQL have the fastest runtime when compared to four other fully-fledged optimizers. Although clearly limited, there are some things to be learned from the MapD approach.

\begin{figure*}[htbp]
	\centering
	\includegraphics[width=.99\textwidth]{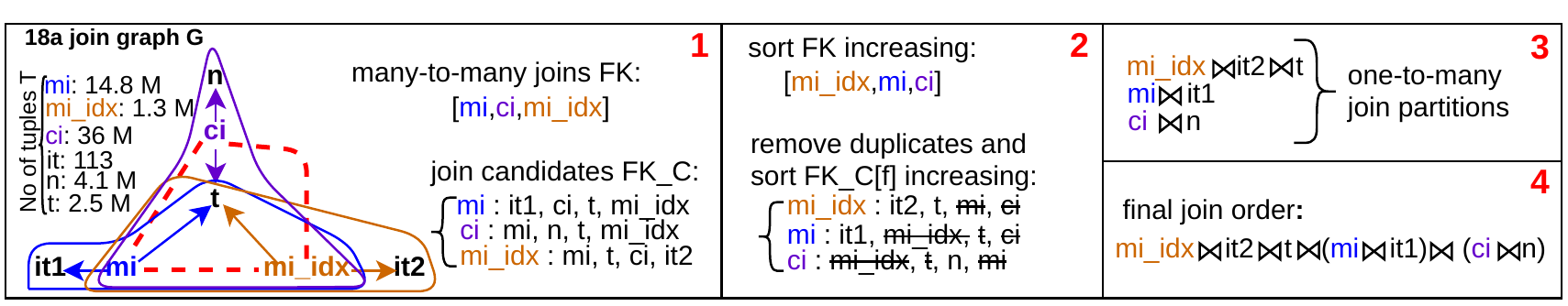}
	\caption{$\text{Simpli}^{2}$ example on JOB query 18a. Key/foreign key constraints are represented by arrows pointing to the key attribute. Many-to-many joins are depicted with dashed lines. Join candidates are grouped into polygons.}
	\label{fig:alg-18a}
\end{figure*}

\smallskip
\noindent
In this paper, we present the Simpli-Squared (or $\text{Simpli}^{2}$) join ordering algorithm that follows the statistics-free approach in MapD. The only other information -- beyond the base table sizes -- taken as input by $\text{Simpli}^{2}$ is the database schema that includes referential integrity constraints or key/foreign key definitions. These are readily available in the database catalog and have been used in previous solutions that improve JOB performance~\cite{Cai:PCETUB:sigmod-2019,Hertzschuch:simplicity:cidr-2021}. $\text{Simpli}^{2}$ computes the join order for a query by splitting the join graph along the many-to-many ($n:m$) joins induced by the referential integrity constraints and sorting the tables based on their size (Figure~\ref{fig:alg-18a}). The tables involved in one-to-many ($1:n$) joins are greedily added to the order based on their size. Cartesian products -- joins between tables that do not share a key -- are avoided throughout the entire process. In summary, $\text{Simpli}^{2}$ is a deterministic algorithm that computes the join order of a query without using any statistics or cardinality estimates. Thus, it incurs negligible overhead.

\smallskip
\noindent
Following prior work~\cite{Hertzschuch:simplicity:cidr-2021}, we implement $\text{Simpli}^{2}$ as a query rewriting procedure that transforms the input query into a series of subqueries corresponding to every split in the join graph. The subqueries are connected together by the cross-split join predicates following the inferred join order. The rewritten query has to be plugged in and executed as is in the target database. This requires disabling the subquery unnesting optimization procedure. Our code is freely available online~\cite{Datta:simpli2-github:2021}.

\smallskip
\noindent
Given its simplicity, what makes $\text{Simpli}^{2}$ relevant are the experimental results on JOB in PostgreSQL---the default target to evaluate query optimization algorithms on. $\text{Simpli}^{2}$ incurs a runtime increase of only up to 16\% -- and in many cases even a reduction -- when compared to four advanced optimizers: the PostgreSQL optimizer, Pessimistic~\cite{Cai:PCETUB:sigmod-2019}, Simplicity~\cite{Hertzschuch:simplicity:cidr-2021}, and COMPASS~\cite{Izenov:compass:sigmod-2021}. All these optimizers use statistics for cardinality estimation and cost functions. The results hold across various configurations, including no indexes, primary key indexes, primary+foreign key indexes, and with/without referential integrity constraints. They also hold when the parameters of the PostgreSQL optimizer cost function are tuned to reflect the relative performance between data access and CPU processing---essentially, tuning PostgreSQL to optimize queries for in-memory processing. Overall, $\text{Simpli}^{2}$ and Pessimistic have the most stable performance across all the settings, while the other optimizers encounter cases where their plans are clearly suboptimal. However, Pessimistic is not practical due to its requirement to build sketches for all the enumerated joins, which incurs a significant overhead. The benefits of $\text{Simpli}^{2}$ are also confirmed on MapD, on which the complete JOB benchmark executes in approximately 200 seconds. With the default optimizer, MapD times out after 10 minutes for 21 queries. The cumulative runtime for the remaining queries is approximately one hour.

\smallskip
\noindent
Two intriguing questions can be raised based on our experimental results. The first question is whether JOB adequately tests query optimizers. While there is no doubt about its complexity, the well-defined structure imposed by key/foreign key relationships -- when appropriately taken into account -- allows for unintended simplifications in join ordering. The importance of cardinality estimation is diminished in this case. Thus, JOB may be more suited for the restricted cardinality estimation task instead of end-to-end query optimization. This leads to the second question: Is cardinality estimation an indispensable requirement in query optimization? Focusing exclusively on devising accurate cardinality estimates does not necessarily result in commensurate query performance improvement. The reason is the complex interaction between cardinality estimation and join enumeration. Since execution time is the definitive query performance metric, these two tasks cannot easily be decoupled. Cardinality estimation has to be considered in the context of join ordering for query optimization---which is not the case in approximate query processing. If we limit ourselves exclusively to JOB, there are not sufficient arguments to include cardinality estimation in query optimization.

\begin{algorithm}[htbp]
	\caption{Simpli-Squared ($\text{Simpli}^{2}$)}\label{alg:the_alg}
	\begin{algorithmic}[1]
		
		\Statex {\bf Input:} query join graph G
		\Statex \hspace{.2cm} database schema S with key/foreign key constraints
		\Statex \hspace{.2cm} number of tuples T in database tables
		\Statex {\bf Output:} join order O, initialized to empty set, $O \leftarrow \emptyset$
		
		\Statex $\triangleright$ \textbf{Split join graph among many-to-many joins}
		
		\State Let FK be the set of tables from G that participate in many-to-many joins according to schema S
		\State Let FK\_C be the join candidates corresponding to every table f in FK (these are the neighbors of f in G)
		
		\Statex $\triangleright$ \textbf{Consider the splits independently}
		
		\While{not all tables in FK are included in O}
		\State Let f be the smallest table in FK [that joins with any other table f' in O]
		
		\Statex $\triangleright$ \textbf{Handle disconnected many-to-many join graph}
		
		\State \textbf{if} no such table f exists 
		\Statex \hspace{.8cm}Let f be the smallest table in G not included in O that joins with any other table f' in O
		\Statex \hspace{.8cm}Insert f in O after the leftmost f'
		\Statex \hspace{.8cm}\textbf{continue}
		\State \textbf{endif}
		
		\Statex $\triangleright$ \textbf{Create one-to-many join partition}
		
		\State Append f to O
		
		\State Remove from FK\_C[f] all tables f' in FK or already included in O
		
		\State Sort FK\_C[f] in ascending order based on T
		
		\State Append FK\_C[f] to O
		
		\State Remove f from FK
		\EndWhile
		
		\Statex $\triangleright$ \textbf{Include one-to-many joins}
		
		\While{not all tables in G are included in O}
		\State Let f be the smallest table in G not included in O [that joins with any other table f' in O]
		\State Insert f in O after the leftmost f'
		\EndWhile
		
	\end{algorithmic}
\end{algorithm}

\section{SIMPLI-SQUARED JOIN ORDERING}\label{sec:simpli-squared}

The design goal of the $\text{Simpli}^{2}$ join ordering algorithm is radical---to not use any sort of synopses/statistics or cardinality estimates. The only inputs -- beyond the query join graph -- are the key/foreign key constraints and the size of the tables. These are used to annotate the vertices (tables) and edges (joins) of the join graph. Tables have their size property, while joins are classified as one-to-many (1:n) or many-to-many (n:m) based on the key/foreign key constraints. $\text{Simpli}^{2}$ treats join ordering as a graph problem where the objective is to determine the ``optimal'' vertex traversal, which is built incrementally starting from a source vertex. Subsequent vertices are greedily added to the traversal based on a set of heuristics aimed at minimizing the size of the current partial ordering. First, there has to be an edge to some already included vertex. This eliminates Cartesian products that always have a larger size than their inputs. Second, one-to-many joins are given priority over many-to-many joins because their size is more predictable---as large as the table with the foreign key. This implies that one-to-many joins are selected first and many-to-many joins are included only when they are necessary to avoid Cartesian products. Finally, the number of tuples is the only measure used to choose between tables, with smaller tables having priority over larger tables. An intuitive explanation for these heuristics is based on the JOB schema. Many-to-many joins typically involve two large fact tables, while one-to-many joins are between one of the fact tables and some (much) smaller dimension table. Prioritizing the one-to-many joins reduces the size of the fact tables by the time they are joined together, while ordering the tables based on their size diminishes the probability that large intermediate results are generated early in the plan.

\smallskip
\noindent
$\text{Simpli}^{2}$ is presented in Algorithm~\ref{alg:the_alg}. It starts with the identification of the tables FK involved in many-to-many joins (line 1) and their join candidates FK\_C (line 2). As depicted in Figure~\ref{fig:alg-18a}, these operations split the join graph into multiple -- possibly overlapping -- components. There is one such component for every table in FK. $\text{Simpli}^{2}$ handles every split independently by creating a one-to-many join partition starting with the table in FK and including its one-to-many join candidates in increasing order of their size (line 7--10). If a table belongs to multiple splits, it is included in the split selected first. The rationale is the smaller size of the FK table. The splits are considered based on the size of the identifying FK table, from small to large (line 4). A special case -- not encountered in the JOB benchmark -- is when the splitting components are disconnected. This prohibits the transition to a new component (line 5). However, since the join graph is connected, there exists a set of tables participating only in one-to-many joins that connect the splits. These tables are greedily inserted in the join ordering immediately after the leftmost table they join with. The same strategy is applied to finalize the join order with orphan one-to-many joins (line 13--16).

\smallskip
\noindent
The execution of $\text{Simpli}^{2}$ on JOB query 18a is depicted in Figure~\ref{fig:alg-18a}. This query has two instances of table \textit{it} and a cycle of many-to-many joins between \textit{ci}, \textit{mi}, and \textit{mi\_idx}. Table \textit{t} is joined with each table in the cycle, creating a clique. This results in a moderately complex join graph for JOB queries. $\text{Simpli}^{2}$ starts by identifying the FK tables and their join candidates. These are all sorted based on the table size. The join partition corresponding to every table in FK is computed by eliminating duplicates. Finally, the partitions are joined together into the final ordering.

\begin{figure*}[htbp]
	\centering
	\includegraphics[width=.99\textwidth]{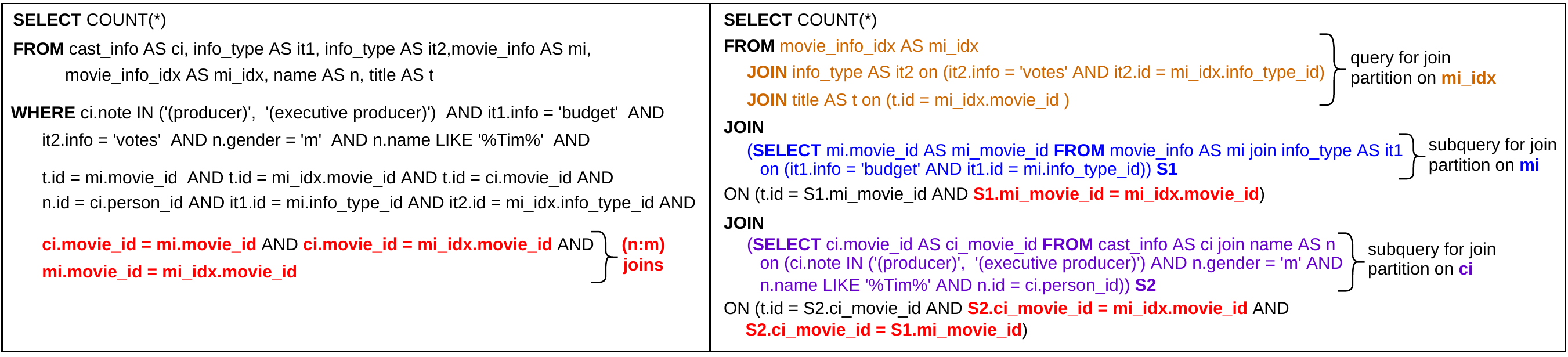}
	\caption{Query rewriting with the explicit join ordering computed by $\text{Simpli}^{2}$ (right) for JOB query 18a (left).}
	\label{fig:rewrite-18a}
\end{figure*}

\smallskip
\noindent
Following the approach in~\cite{Hertzschuch:simplicity:cidr-2021}, we show in Figure~\ref{fig:rewrite-18a} how the original query is rewritten to use the explicit join ordering. This involves the creation of a subquery for every join partition -- except the first one -- and the renaming of the join attributes used outside the subquery. When the explicit query is executed inside a database, e.g., PostgreSQL or MapD, the optimizer has to be instructed to follow the exact join order and subquery unnesting has to be disabled. Nonetheless, the optimizer still has to select the physical operators. This process may use statistics if multiple alternatives exist, which is the case when indexes are available.

\section{JOIN ORDERING WITH CARDINALITY ESTIMATES}\label{sec:join-order-card-est}

In this section, we present the standard cardinality estimation approach that uses a variety of data synopses and statistics for join ordering. First, we introduce the PostgreSQL implementation and then discuss three recently proposed methods. Two of these methods -- Pessimistic~\cite{Cai:PCETUB:sigmod-2019} and Simplicity~\cite{Hertzschuch:simplicity:cidr-2021} -- combine estimates and the structure of the join graph to determine the order, while COMPASS~\cite{Izenov:compass:sigmod-2021} uses statistics exclusively.

\paragraph*{PostgreSQL}
The PostgreSQL query optimizer enumerates all possible join orderings -- or plans -- exhaustively to find the one with the minimum cost. While the enumeration can benefit from the join graph by traversing only existing edges, the type of the edge is not a primary parameter. Redundant and suboptimal orderings are efficiently handled through dynamic programming and early pruning. The cost of a plan is computed by aggregating the cost of the individual plan operators. The cost of an operator depends on the number of accessed pages and the number of processed tuples. The weight of these terms is controlled by configurable system parameters. The cost terms are known only for the base tables, and they have to be estimated for all the other operators without actually executing them. This is the cardinality estimation problem. Pre-computed data synopses or statistics are used for this purpose. PostgreSQL statistics are attribute-level and include the range, the most common values or heavy hitters, the number of distinct values, and an equi-depth histogram. The cost of an operator is estimated by combining these statistics into formulas that make uniformity, independence, and inclusion assumptions about the data they are derived from. This results in inaccurate estimates -- thus, an incorrect plan cost -- whenever the assumptions do not hold, such as when there are many-to-many joins and long sequences of selections and joins. In summary, cardinality estimates are the primary factor in computing the optimal join ordering, while the structure of the join graph and the join type are mostly ignored.

\begin{table*}[htbp]
	\centering
	\begin{adjustbox}{width=1\textwidth}
		\begin{tabular}{|l||llll|}
			
			\hline
			\textbf{Method} & \textbf{Statistics} & \textbf{Estimates} & \textbf{Join Graph Use} & \textbf{PostgreSQL Integration} \\
			
			\hline
			PostgreSQL & distincts, heavy hitters, histograms & all & minimal & native \\
			Pessimistic & Count-Min sketches, PostgreSQL & many-to-many joins & join type & estimates in optimizer \\
			Simplicity & table sizes, heavy hitters, PostgreSQL & all & join type & query rewriting \\
			COMPASS & Fast-AGMS sketches & joins & graph & query rewriting \\
			
			\textbf{$\text{Simpli}^{2}$} & \textbf{table sizes} & \textbf{none} & \textbf{join type} & \textbf{query rewriting} \\
			\hline
			
		\end{tabular}
	\end{adjustbox}
	\caption{Analytical comparison between $\text{Simpli}^{2}$ and cardinality estimates approaches.}
	\label{tbl:analytic-comp}
\end{table*}

\paragraph*{Pessimistic}
The main idea of the Pessimistic approach proposed in~\cite{Cai:PCETUB:sigmod-2019} is to use upper bounds instead of estimates for the cardinality of difficult many-to-many multi-way joins. The upper bounds are derived with multidimensional Count-Min sketches computed online over the result of one-to-many joins and selections. In this way, the only estimates used in join ordering are for many-to-many joins -- the other cardinalities are exact -- significantly reducing the chance for inaccurate plan costs. There are two limitations to this approach. First, the size of the multidimensional sketches increases exponentially with the number of joins. For this reason, they are kept 3D at most---which is not an issue for JOB. Second, the overhead to compute the sketches at query time is substantial~\cite{Hertzschuch:simplicity:cidr-2021}, rendering it impractical. Overall, Pessimistic reduces the number of cardinality estimates used in join ordering only to many-to-many joins while keeping the other components of the PostgreSQL optimizer unmodified.

\paragraph*{Simplicity}
To address the shortcomings of Pessimistic, the Simplicity solution~\cite{Hertzschuch:simplicity:cidr-2021} replaces Count-Min sketches with the already available PostgreSQL statistics in order to compute the upper bounds. Unlike Pessimistic, though, upper bounds are computed for all the joins, including one-to-many. However, the one-to-many joins are still given priority based on the join graph, while many-to-many joins are only performed when necessary. Simplicity generates the join ordering outside of the PostgreSQL optimizer and enforces it through query rewriting. Thus, while upper bounds are used to determine the ordering, the standard PostgreSQL estimates are still used to select the physical operators.

\paragraph*{COMPASS}
Fast-AGMS sketches are the statistics used for join cardinality estimation in COMPASS~\cite{Izenov:compass:sigmod-2021}. Fast-AGMS sketches give unbiased estimates, while Count-Min always overestimates---thus, the upper bounds in Pessimistic. In COMPASS, sketches are computed on the base tables -- not for every enumerated join -- which allows for incremental merging in order to estimate any join sequence. This is a different solution from Simplicity to handle the practicality issue in Pessimistic. COMPASS uses a greedy approach to determine the join order, which is always a left-deep tree. The join type is not considered at all. While the Fast-AGMS estimates can be directly embedded in the PostgreSQL optimizer, the query plans are integrated through rewriting.

\paragraph*{Summary}
We classify $\text{Simpli}^{2}$ and the presented methods based on the extent they use statistics for cardinality estimation, the use of the join graph in computing the ordering, and the integration in PostgreSQL (Table~\ref{tbl:analytic-comp}). $\text{Simpli}^{2}$ does not use any statistics nor cardinality estimates. Instead, it uses only table sizes. Pessimistic and COMPASS derive estimates based exclusively on correlated sketches, while PostgreSQL and Simplicity use multiple types of statistics. While PostgreSQL, Simplicity, and COMPASS use estimates for all the joins, Pessimistic estimates only many-to-many joins---it uses exact cardinality for one-to-many joins. The join graph and/or join type are used the least in PostgreSQL and COMPASS. Pessimistic, Simplicity, and $\text{Simpli}^{2}$ prioritize one-to-many joins and delay the inclusion of many-to-many joins to when absolutely necessary. Finally, Pessimistic is integrated with the PostgreSQL optimizer by directly plugging in its bounds. The other solutions use query rewriting to enforce their join ordering. This, however, allows PostgreSQL to use its own statistics to select the physical operators. To summarize, the only significant difference between $\text{Simpli}^{2}$ and the other methods is the lack of cardinality estimates and a cost model.

\section{EXPERIMENTAL EVALUATION}\label{sec:experiments}

We perform an extensive experimental study that compares $\text{Simpli}^{2}$ with all the statistics-based methods included in Table~\ref{tbl:analytic-comp} on the complete JOB benchmark. Given that it is estimation-free, the expectation is that $\text{Simpli}^{2}$ is worse. Nevertheless, our experiments find the difference to only be up to 16\% of the best method across all the settings. This result is notable, considering the minimalistic requirements $\text{Simpli}^{2}$ has compared to the other methods. Aside from this, the questions we investigate are:

\begin{itemize}[leftmargin=*,noitemsep]
\item What is the execution time -- or runtime -- of the considered methods in PostgreSQL? Is it consistent in the presence of no indexes (No Index), primary key indexes (PK), and primary+foreign key indexes (PK+FK)?

\item How do the methods perform when we tune the PostgreSQL optimizer cost function to reflect the relative performance between data access and CPU processing? By default, this relative ratio is 100, which means that a page data access takes 100 times more than the CPU processing of a tuple. We revisit this assumption and consider two other configurations in which the ratio is reduced to 10 and 1, respectively. Surprisingly, this simple change gives some of the methods a significant boost.

\item Does $\text{Simpli}^{2}$ work for modern memory-centric databases? We confirm this on MapD and MonetDB. While $\text{Simpli}^{2}$ improves tremendously over MapD due to its simplistic optimizer, surprisingly, it also improves over the more advanced MonetDB.

\item Is the $\text{Simpli}^{2}$ performance consistent across different query workloads? Experimental results on the JOB-light workload -- derived from JOB -- confirm that $\text{Simpli}^{2}$ achieves comparable performance with the other methods.
\end{itemize}

\subsection{Setup}\label{ssec:setup}

\paragraph*{Implementation}
We implement $\text{Simpli}^{2}$ as a Python script -- inspired from the Simplicity code~\cite{Hertzschuch:simplicity-github-repo:2020} -- consisting of two modules. The first module implements join reordering (Algorithm~\ref{alg:the_alg}), while the second is a lightweight query rewriting procedure that transforms the original SQL statement into the explicit form suggested by the reordering (Figure~\ref{fig:rewrite-18a}). The source code is publicly available~\cite{Datta:simpli2-github:2021}.

\paragraph*{Hardware and database}
We run the experiments in a Ubuntu 20.04 LTS docker container on a machine with 56 CPU cores (Intel Xeon E5-2660), 256 GB RAM, and HDD storage. We use PostgreSQL 14.2 configured with 128 GB memory limit per operator (\texttt{work\_mem}), 128 GB buffer pool size (\texttt{shared\_buffers}), and 128 GB OS buffer cache (\texttt{effective\_} \texttt{cache\_size}). We also increase the \texttt{geqo\_threshold} parameter to 18 joins, which determines the threshold for switching from dynamic programming to heuristic search. Lastly, we change \texttt{from\_collapse\_limit} and \texttt{join\_collapse\_limit} to 1 for Simplicity, COMPASS, and $\text{Simpli}^{2}$, so that the query optimizer does not reorder joins or unnest subqueries.

\begin{figure*}[htbp]
	\centering
	\begin{subfigure}{.8\textwidth}
		\includegraphics[width=\textwidth]{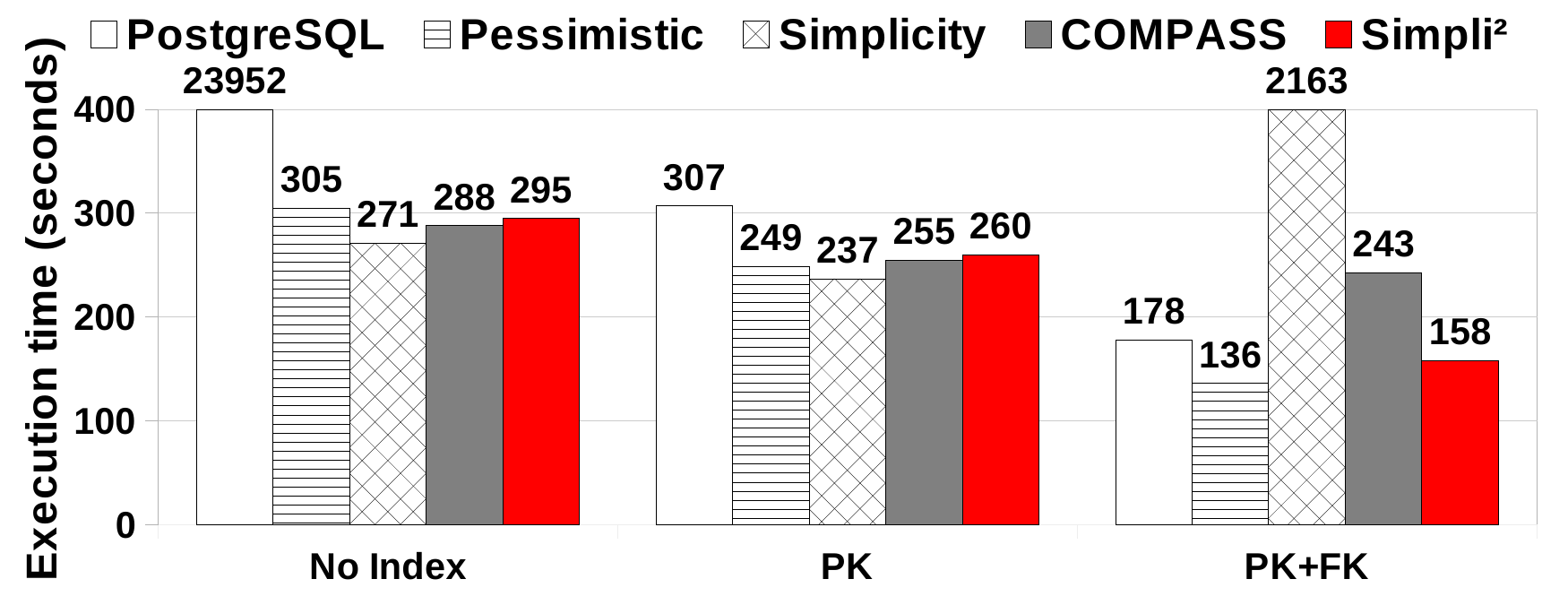}
		\caption{data/cpu = 100}
		\label{fig:cpu.01}
	\end{subfigure}

	\vspace*{1cm}

	\begin{subfigure}{.8\textwidth}
		\includegraphics[width=\textwidth]{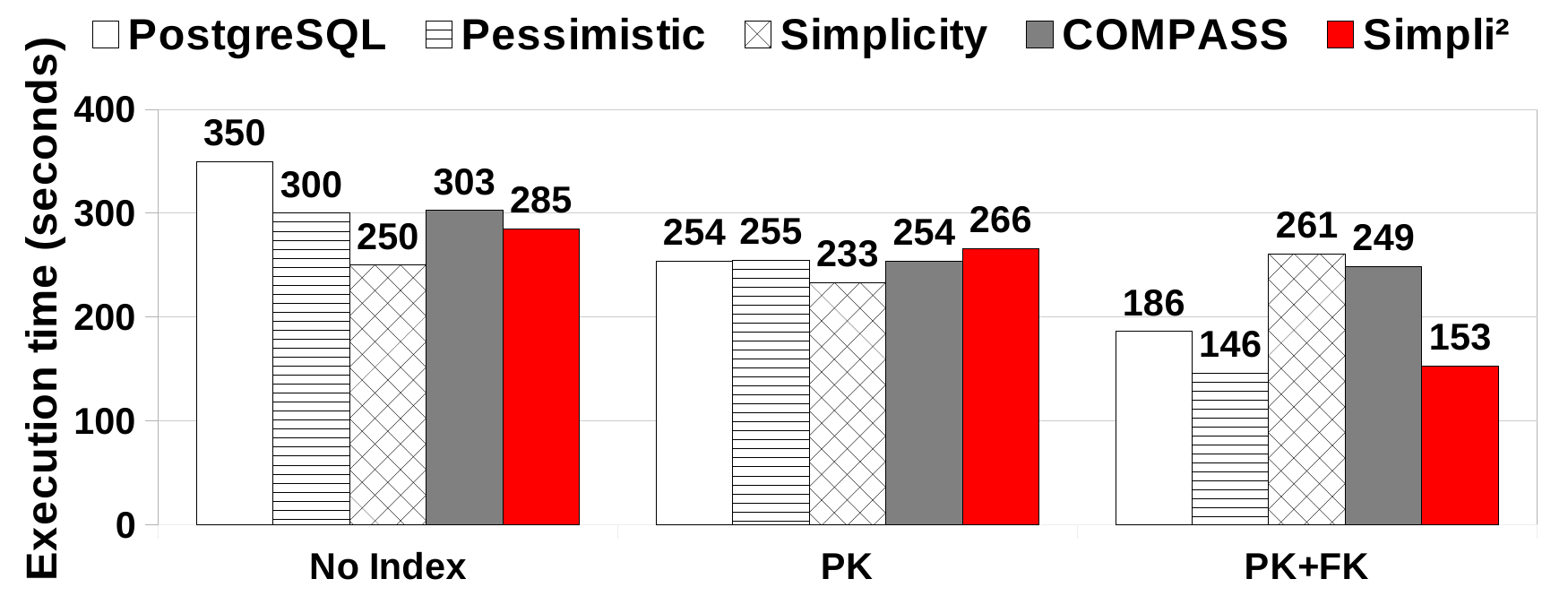}
		\caption{data/cpu = 10}
		\label{fig:cpu.1}
	\end{subfigure}

	\vspace*{1cm}

	\begin{subfigure}{.8\textwidth}
		\includegraphics[width=\textwidth]{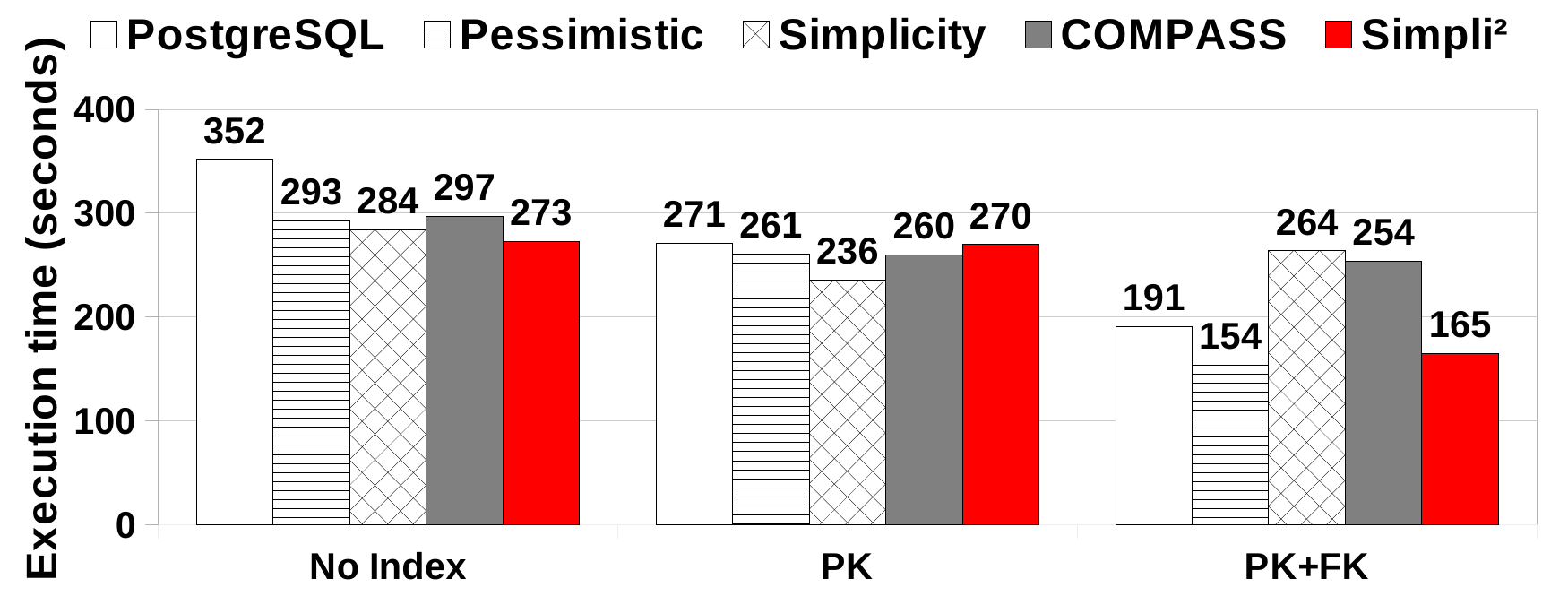}
		\caption{data/cpu = 1}
		\label{fig:cpu1}
	\end{subfigure}
	\caption{JOB execution time in PostgreSQL as a function of the ratio between data access and CPU processing.}
	\label{fig:job-runtime}
\end{figure*}

\paragraph*{Methodology}
We use the total runtime of the JOB workload and the maximum runtime of a query as our evaluation metrics. These times include only the query execution. They do not include the sketch building time for Pessimistic and COMPASS. The reported results are the median over 5 runs. In order to fairly evaluate the different join orders, we use PostgreSQL as common ground. We run all the methods independently and collect their join orders and explicit query statements. Then, we inject these statements in PostgreSQL and measure their runtime. For Pessimistic, we port the modification in the cost function~\cite{Cai:pessimistic-github:2019} to PostgreSQL 14.2 and run it as a standalone instance.

\smallskip
\noindent
The combination of indexes and ratio between data access and CPU processing -- we experiment with the ratio because our hardware can cache the entire database -- gives 9 configurations for every method. We run each of these configurations with and without referential integrity constraints in the schema declaration. We report only the best result, which varies among configurations.

\subsection{Results}\label{sssec:results}

\paragraph*{JOB Benchmark}
Figure~\ref{fig:job-runtime} depicts the JOB workload runtime, while Table~\ref{tbl:max-query-time} contains the maximum runtime across all the queries. In the following, we analyze the results organized by methods.

\smallskip
\noindent
\textit{PostgreSQL.}
With the default ratio data/cpu of 100 and no indexes, PostgreSQL has more than 10 queries that take over 100 seconds to execute---including an outlier that takes 7690 seconds. Tuning the ratio to the hardware characteristics has a dramatic impact, reducing the JOB runtime to merely 350 seconds---which is in line with the other methods. The ratio impacts the choice of the physical operators, prioritizing hash join. With the addition of indexes, the runtime is further decreased to as low as 178 seconds for PK+FK. The ratio does not help much for the overall runtime. However, it reduces the maximum runtime of a query with PK to 6.75 seconds---the smallest for this configuration type. Overall, a ratio of 10 generates the best results, allowing PostgreSQL to perform comparably to most of the other methods.

\smallskip
\noindent
\textit{Pessimistic.}
Pessimistic with PK+FK indexes achieves the fastest runtime across all the configurations and methods. The characteristic of PK+FK plans is the extensive use of nested loop joins with indexed scans. The exact cardinalities for one-to-many joins and the upper bound estimates allow the optimizer to choose the best plan---both join order and physical operators. However, the more accurate cardinality estimates do not seem to help much for configurations with fewer or no indexes. The caveat of this method remains the sketch building time, which -- if included -- significantly increases the time for queries with 6 or more joins~\cite{Hertzschuch:simplicity:cidr-2021}.

\begin{table*}[htbp]
	\centering
	\begin{tabular}{|l||rrr|rrr|rrr|}
		
		\hline
		\multicolumn{1}{|c||}{\multirow{3}{*}{\textbf{Method}}} & \multicolumn{9}{c|}{\textbf{data/cpu}} \\
		\multicolumn{1}{|c||}{} & \multicolumn{3}{c|}{\textbf{100}} & \multicolumn{3}{c|}{\textbf{10}} & \multicolumn{3}{c|}{\textbf{1}} \\
		\multicolumn{1}{|c||}{} & \textbf{No Index} & \textbf{PK} & \textbf{PK+FK} & \textbf{No Index} & \textbf{PK} & \textbf{PK+FK} & \textbf{No Index} & \textbf{PK} & \textbf{PK+FK} \\
		\hline
		
		PostgreSQL & \textbf{7690} & \textbf{45.29} & 14.71 & \textbf{31.53} & 6.75 & 16.56 & \textbf{32.96} & 9.40 & 16.76 \\
		Pessimistic & 12.92 & 11.88 & 10.36 & 13.52 & \textbf{12.65} & 13.04 & 12.33 & \textbf{12.53} & 11.69 \\
		Simplicity & 6.17 & 7.47 & \textbf{751} & 5.50 & 7.44 & \textbf{60.45} & 12.41 & 7.48 & \textbf{63.64} \\
		COMPASS & 16.31 & 7.68 & 33.37 & 29.83 & 8.21 & 32.62 & 29.86 & 7.89 & 33.10 \\
		\textbf{$\text{Simpli}^{2}$} & 11.07 & 10.60 & 10.72 & 10.85 & 10.15 & 10.06 & 10.55 & 10.75 & 10.26 \\
		\hline
		
	\end{tabular}
	\caption{The maximum execution time (seconds) for a query in PostgreSQL.}
	\label{tbl:max-query-time}
\end{table*}

\smallskip
\noindent
\textit{Simplicity.}
When none or only PK indexes are available, Simplicity is the clear winner---both in terms of overall and maximum runtime. This is due to the combination of statistics and key/foreign key constraints utilized to identify the optimal join ordering. When FK indexes are also available, the situation changes dramatically. A modification of the ratio parameter is required to bring Simplicity to acceptable levels---still much larger than Pessimistic. The reason for this behavior is the use of query rewriting to enforce join ordering. The use of unnested subqueries limits the applicability of index scans to the subquery. The joins between subqueries require a different type of join. When a limited set of indexes are available, the choice is simple---hash join. With indexes, the decision becomes complicated and prone to errors. The ratio parameter steers the optimizer toward hash joins, similar to how it is restricted in~\cite{Hertzschuch:simplicity:cidr-2021}.

\smallskip
\noindent
\textit{COMPASS.}
Although it uses sketches exclusively, COMPASS achieves similar results to the other methods when only limited indexes exist. Its performance does not improve in the PK+FK case because the selected join order does not account for index scans. However, since the plans are always left-deep without subqueries, COMPASS avoids the bad decisions made by Simplicity.

\smallskip
\noindent
\textit{$\text{Simpli}^{2}$.}
Given its statistics-free nature, $\text{Simpli}^{2}$ achieves surprisingly good results. It is always within 16\% of the fastest method on the complete JOB, and its maximum runtime for a query is never the worst. This confirms the intuition $\text{Simpli}^{2}$ is built on. With limited indexes, the preferred join operator is hash join. Since all the methods rely on it, there is a limited improvement margin accurate estimates can bring on top of wisely using the join graph and type---which $\text{Simpli}^{2}$ does. PK+FK indexes improve the overall runtime significantly if suitable physical operators are selected. To avoid the limitations of query rewriting, $\text{Simpli}^{2}$ considers two alternatives for the join ordering it finds---with and without subqueries. The plan with subqueries is identical to the limited indexes case. The main drawback is that index scans are restricted to subqueries. The plan without subqueries is a flat left-deep tree in which index scans are more frequently used. Since the best alternative varies across queries, we report the fastest runtime---which can readily be found by running the two choices concurrently.

\begin{table}[htbp]
	\centering
	\begin{tabular}{|l||rrr|}

		\hline
		\textbf{Method} & \textbf{No Index} & \textbf{PK} & \textbf{PK+FK} \\
		\hline

		MonetDB & 44.60 (2.56) & 44.13 (2.54) & 45.73 (2.53) \\
		Simplicity & 38.86 (2.51) & \textbf{31.86} (\textbf{2.41}) & 38.04 (2.49) \\
		\textbf{$\text{Simpli}^{2}$} & \textbf{35.20} (\textbf{1.84}) & 37.98 (2.49) & \textbf{37.70} (\textbf{1.93}) \\
		\hline

	\end{tabular}
	\caption{Total (maximum per query in parentheses) execution time (seconds) in MonetDB.}
	\label{tbl:monetdb}
\end{table}

\paragraph*{MonetDB}
For completeness, we report results on MonetDB 11.41.5~\cite{monetdb-web}, although the exact execution of a join ordering is not guaranteed in this case. In this comparison, we include only the native MonetDB optimizer, Simplicity, and $\text{Simpli}^{2}$ since results have been published previously for the former two. The results are included in Table~\ref{tbl:monetdb}. Both Simplicity and $\text{Simpli}^{2}$ improve upon native MonetDB in all the configurations. While Simplicity PK has the fastest overall runtime, surprisingly, $\text{Simpli}^{2}$ is fastest without indexes and with PK+FK. As far as we know, statistics do not play such an important role in MonetDB, thus, the success of $\text{Simpli}^{2}$. Moreover, indexes do not seem to help much with any of the methods---except Simplicity PK.

\begin{table}[htbp]
	\centering
	\begin{tabular}{|l||rrr|}

		\hline
		\textbf{Method} & \textbf{No Index} & \textbf{PK} & \textbf{PK+FK} \\
		\hline

		PostgreSQL & 849 (\textbf{291}) & 851 (\textbf{291}) & 777 (\textbf{222}) \\
		Pessimistic & \textbf{747} (299) & \textbf{754} (294) & \textbf{690} (237) \\
		\textbf{$\text{Simpli}^{2}$} & 882 (312) & 891 (311) & 789 (231) \\
		\hline

	\end{tabular}
	\caption{Execution time (seconds) for the JOB-light benchmark in PostgreSQL. The numbers in parentheses correspond to the execution time obtained when two outlier queries -- which take more than all the other 68 queries combined -- are excluded. The outlier queries are the 58th and 60th in the workload.}
	\label{tbl:job-light}
\end{table}

\paragraph*{JOB-light Benchmark}
JOB-light is a workload derived from JOB that contains 70 queries with a maximum of six tables in its most complex queries. The majority of the join graphs in JOB-light form a star-join structure. JOB-light does not contain complex selection predicates, such as \texttt{LIKE} on strings and disjunctions. In fact, the selection predicates are exclusively applied to numerical and categorical attributes. However, the simplified JOB-light is frequently utilized for evaluating learning-based cardinality estimation algorithms~\cite{Kipf:LCECJDL:cidr-2019}. This is the main reason $\text{Simpli}^{2}$ is also evaluated on the JOB-light benchmark. The results are included in Table~\ref{tbl:job-light}. A noticeable fact is that the execution time of two complex queries dominates the overall workload due to the nonexistence of direct join relations between the tables. There are no key/foreign key relationships, which leads to cross joins. Pessimistic~\cite{Cai:PCETUB:sigmod-2019} achieves a better result than the other two methods on these two queries because it includes estimates for transitive rules, i.e., if we have two join predicates \texttt{a=b} and \texttt{b=c}, then estimates are also computed for \texttt{a=c}. Except these two queries, all the methods produce similar results, which are within 10\% of each other.

\subsection{Physical Operators Analysis}

The choice of the physical operators impacts the query execution time significantly. For example, choosing merge join over hash join may be beneficial in cases where the joined relations are small and sorted since there is no need to allocate auxiliary memory and move the data to storage. However, by using explicit hints to force a particular join order, independent choices for physical operators are not possible anymore. In the experiments, we observe that the query optimization methods that select a join order while considering the available physical operators and indexes improve execution time compared with methods that produce identical plans through subquery rewriting. We provide graphical evidence for the reasons this happens based on the execution plans for JOB query 18a.

\begin{figure*}[htbp]
	\centering
	\includegraphics[width=.99\textwidth]{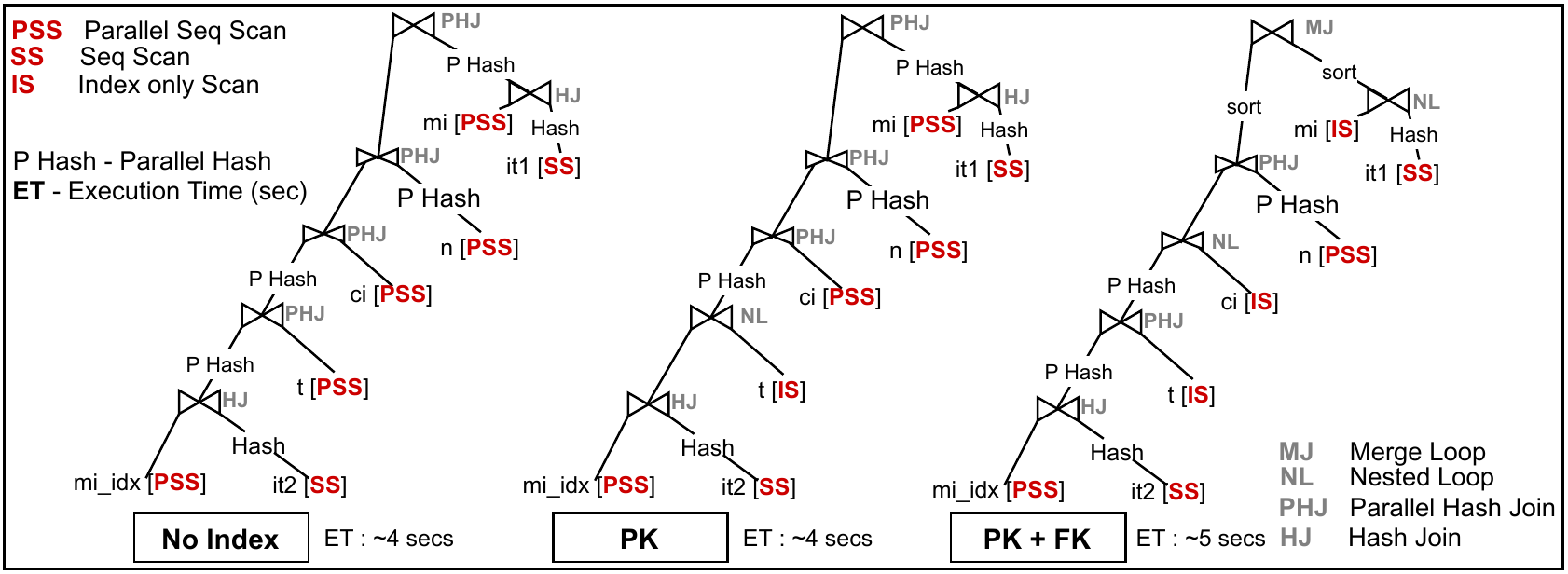}
	\caption{Simplicity: Physical operators in different index configurations.}
	\label{fig:simplicity-index}
\end{figure*}

\smallskip
\noindent
Forced to choose the same join order across different index settings and focusing on bushy join graph structures limits the plan search space in Simplicity~\cite{Hertzschuch:simplicity:cidr-2021}. Therefore, it is challenging to improve the execution time when indexes are available. In Figure~\ref{fig:simplicity-index}, sequential scan is chosen for base table access and hash join as the only join operator in the No Index setting. In the presence of primary key indexes, the base table accesses are still dominated by sequential scan operators, except in the case when an index scan is used for table t, followed by a nested loop join. The other operator choices remain the same. The execution time remains largely the same as there is no significant change in selecting the physical operators. In the PK+FK setting, index scan followed by nested loop join is used for multiple base tables. Also, sort merge join appears in one case. However, all these changes do not benefit the execution time because of the imposed join graph structure. In the PK+FK setting, the query takes 5 seconds to execute, which is worse than without any indexes. This proves that changes in the physical operators within a confined join graph structure provide only limited benefits---if at all.

\begin{figure*}[htbp]
	\centering
	\includegraphics[width=.99\textwidth]{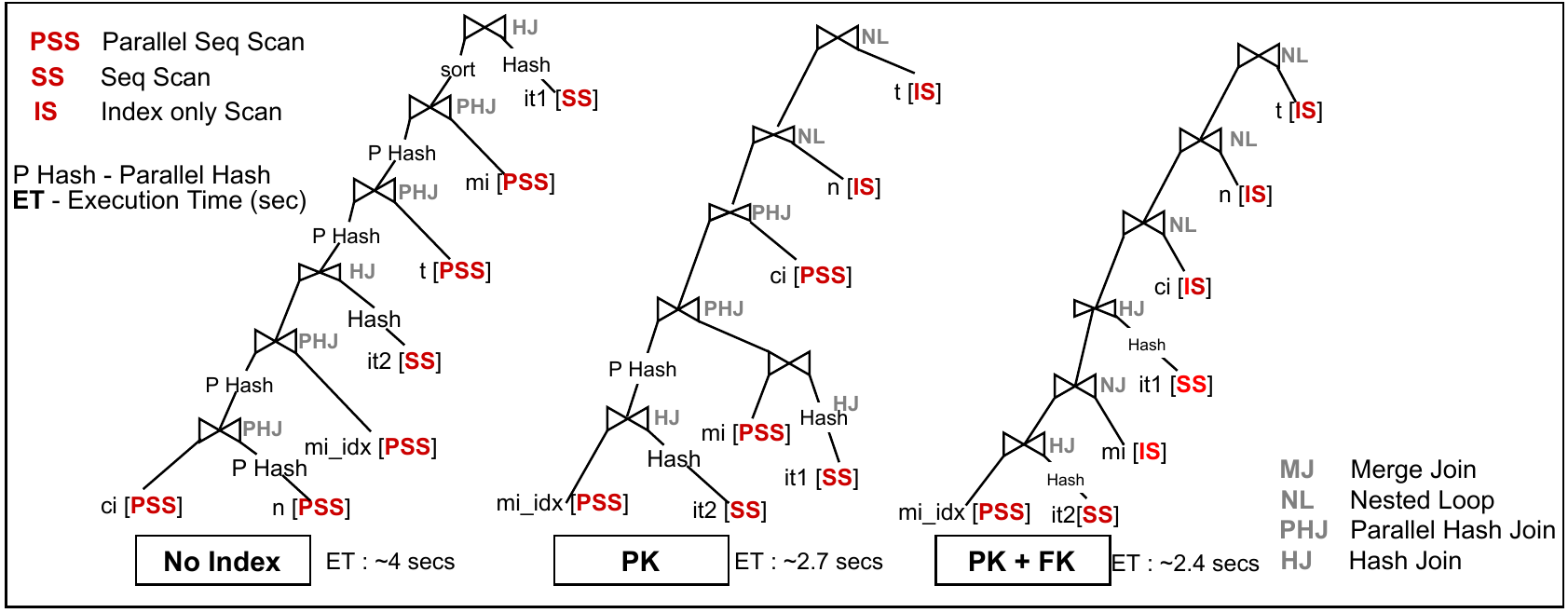}
	\caption{PostgreSQL: Physical operators in different index configurations.}
	\label{fig:postgresql-index}
\end{figure*}

\begin{figure*}[htbp]
	\centering
	\includegraphics[width=.99\textwidth]{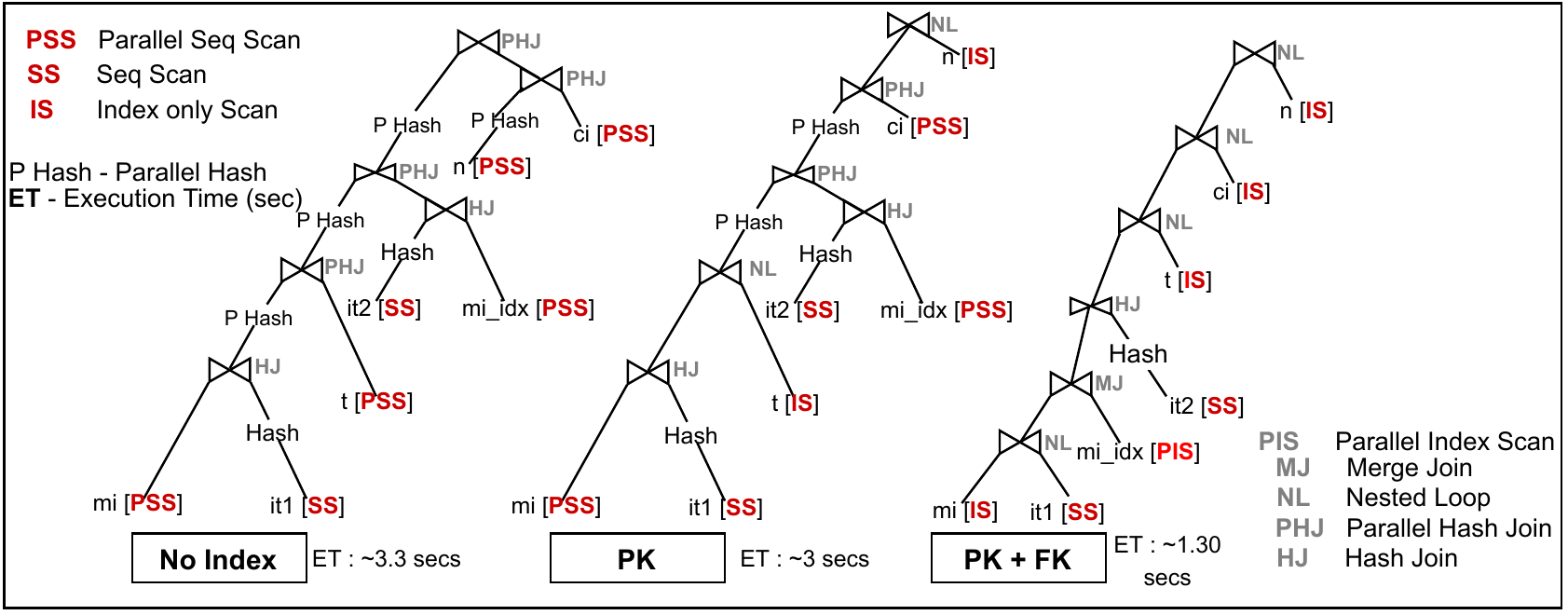}
	\caption{Pessimistic: Physical operators in different index configurations.}
	\label{fig:pessimistic-index}
\end{figure*}

\begin{figure*}[htbp]
	\centering
	\includegraphics[width=.99\textwidth]{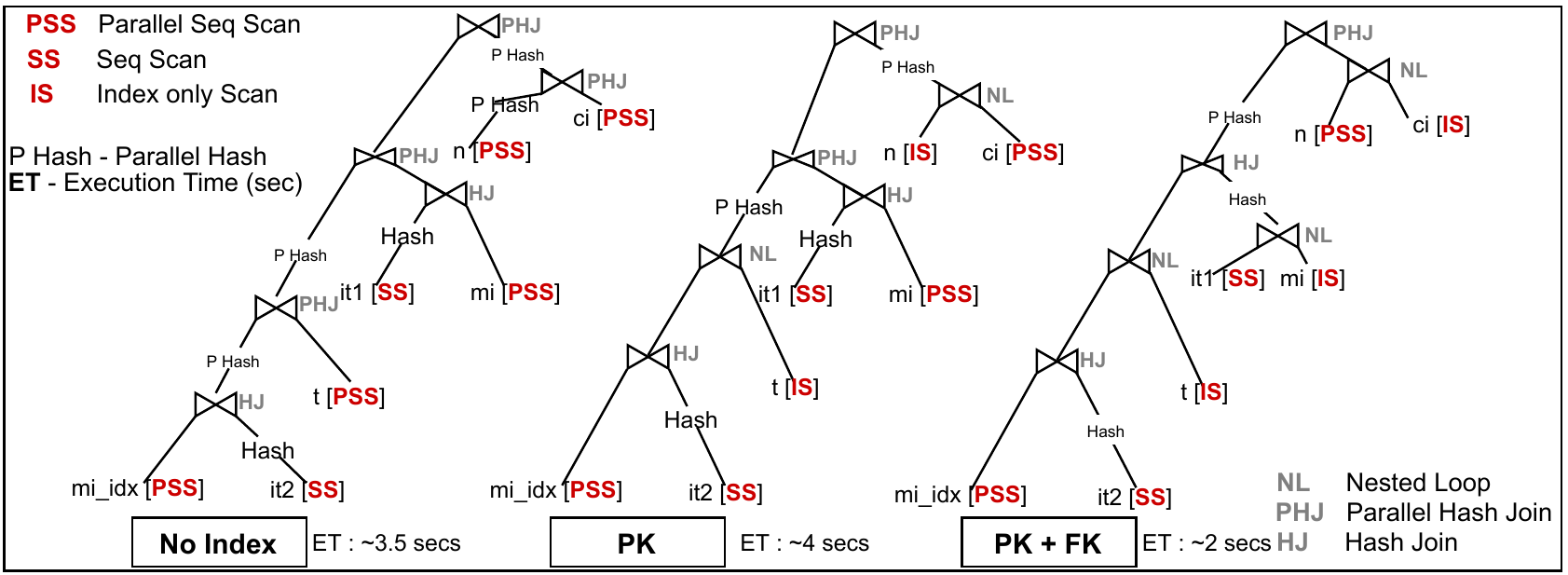}
	\caption{$\text{Simpli}^{2}$: Physical operators in different index configurations.}
	\label{fig:simpli2-index}
\end{figure*}

\smallskip
\noindent
On the contrary, in PostgreSQL, Pessimistic, and $\text{Simpli}^{2}$ the join order changes in the presence of different index configurations in order to take advantage of the optimal physical operators. We observe an improvement in the runtime in most cases due to the changes in the join order and efficient use of the available indexes. In Figure~\ref{fig:pessimistic-index}, Pessimistic chooses a bushy join order plan, including two sub-trees in the No Index setting. Sequential scans and hash joins are used for base table access and join operations, respectively. In the PK setting, the join tree maintains its bushy structure, but includes a single sub-tree in the plan. Index scans are used for the base tables wherever possible. The query execution time is slightly better. In the PK+FK setting, the join graph morphs into a left deep tree, enabling the query to benefit from all the available indexes. In Figure~\ref{fig:postgresql-index} and~\ref{fig:simpli2-index}, PostgreSQL and $\text{Simpli}^{2}$ show a similar trend as Pessimistic across the different index configurations.

\smallskip
\noindent
Based on these diagrams, we can characterize the impact of indexes on the selection of the physical operators. When no indexes are available, the access path to the base tables is exclusively the sequential scan. Hash joins are almost always preferred over sort merge join, while nested loop joins are completely prohibited. Given the limited type of operators, the join order plays the dominant role in determining the execution time. When indexes are available, sequential scans are replaced with indexed scans for base table access. Moreover, indexes allow for direct access based on the join key, thus, they eliminate the need to build hash tables. This is reflected by the replacement of hash joins with index nested loop joins. However, for this to be possible, base tables have to be added to the join graph one-at-a-time. This is reflected in the left-deep tree structure of the execution plan. If the join order is imposed through subqueries, the applicability of indexes is limited. The execution plan cannot transform based on the available indexes. This is the main problem Simplicity faces. This is also the reason $\text{Simpli}^{2}$ considers both the fixed subquery plan and the left-deep tree plan when indexes are available.

\section{RELATED WORK}\label{sec:related-work}   

Finding an optimal join order is the most important decision to be made by a query optimizer. The problem has been one of the most studied tasks in the database field for over four decades. There are two different types of query optimizers, depending on how the join order is determined---\textit{cardinality estimation} and \textit{rule-based} optimizers. 

\paragraph*{Cardinality Estimation Query Optimization}
Cardinality estimation has been an active research direction among database practitioners in the last decade due to its importance in finding optimal query execution plans. Even though Leis et al.~\cite{Leis:JOB:vldb-2018,Leis:QOREALLY:pvldb-2015} show the impact of cardinality estimates in query optimization experimentally, estimations are often inaccurate due to simplified assumptions such as uniform distribution of data, independence, and inclusion. PostgreSQL~\cite{postgres} uses histograms for data representation and formulas based on simplified assumptions for cardinality estimation, which often leads to inaccurate estimates and suboptimal query execution plans. Histograms are accurate for single attribute estimations. However, it is difficult for them to capture join-crossing correlations. Cai et al.~\cite{Cai:PCETUB:sigmod-2019} introduce Pessimistic~\cite{Cai:PCETUB:sigmod-2019}, which uses Count-Min sketches for capturing join crossing correlations. The sketch building process introduces significant overhead when the number of join increases. Hertzschuch et al.~\cite{Hertzschuch:simplicity:cidr-2021} maintain the pessimistic property for cardinality estimation while replacing sketches with a simple formula based on statistics already available to the PostgreSQL query optimizer. Izenov et al.~\cite{Izenov:compass:sigmod-2021} use Fast-AGMS sketches to capture join-crossing correlations, which introduces negligible overhead in the sketch building process compared with Pessimistic. $\text{Simpli}^{2}$ does not require any cardinality estimates for finding a query execution plan, unlike the other methods.

\paragraph*{Rule-based Query Optimization}
Unlike a cardinality estimation-based optimizer, a rule-based optimizer relies on a set of predefined rules to prune the optimization search space and decide on an optimal plan for query execution. Most rule-based optimizers~\cite{ingres,Starburst,orcl-qop,Volcano,EXODUS,Praire,venusdb,calcite,presto,cockroachdb} develop specialized rule languages and execution environments to avoid unwanted compatibility issues and accelerate the enumeration process and query execution. Held et al.~\cite{ingres} introduce the first rule-based optimizer in Ingres, where the original query is decomposed into single-valued subqueries and executed separately in a greedy cardinality-based fashion. While this works well for simple queries, it becomes nearly impossible to obtain a good plan in complex interdependent queries. Unlike Ingres, Starburst, developed by Pirahesh et al.~\cite{Starburst}, is a Query Graph Model (QGM) based optimizer where a SQL query is represented as a graph. Query rewriting rules are used to transform a QGM into another equivalent QGM. Each equivalent QGM is assigned an estimated cost, calculated using optimization rules, in the plan optimization phase. The QGM with minimum cost is chosen for query execution. Graefe et al.~\cite{EXODUS} introduce EXODUS, where a query is represented as an algebraic tree. EXODUS follows rule-based reordering and plan optimization techniques similar to Starburst. A redundant and straightforward search strategy and the simple cost function introduces significant limitations in Starburst and EXODUS, respectively. Graefe et al.~\cite{Volcano} introduce Volcano to address these limitations. Volcano uses directed dynamic search instead of rules for enumeration. Unlike any of these optimizers, $\text{Simpli}^{2}$ is generic and does not rely on any system-specific rules. It uses the query join graph, integrity constraints, and base table cardinality to determine the execution plan.

\paragraph*{Machine Learning Query Optimization}
More recently, machine learning models and deep neural networks have become a topic of interest in query optimization. Many machine learning solutions target individual query optimization components such as cardinality estimation~\cite{Kipf:LCECJDL:cidr-2019,Marcus:NEO:pvldb-2019,Woltmann:CEL:aiDM-2019,
Liu:CEUNN:cascon-2015,Ortiz:EADLCE:arxiv-2019,Malik:BBAQCE:cidr-2007}, but much fewer focus on more overarching problems like join ordering~\cite{Krishnan:LOJQDRL:arxiv-2018,Marcus:DRLJOE:aiDM-2018} or end-to-end query optimization~\cite{Marcus:NEO:pvldb-2019}. Although so-called learned cardinality estimators may be extremely accurate on single table queries~\cite{wang2021ready}, Han et al.~\cite{han2021cardinality} show that these methods may fail to improve end-to-end query execution time, especially when applied to multi-table queries. For join-ordering, reinforcement learning is of particular interest. Krishnan et al.~\cite{Krishnan:LOJQDRL:arxiv-2018} propose using deep reinforcement learning to learn a value model that replaces traditional heuristics in guiding plan enumeration. Marcus et al. propose ReJOIN~\cite{Marcus:DRLJOE:aiDM-2018}, which also uses deep reinforcement learning, but to learn a policy model instead of a value model. There are trade-offs between policy optimization and value learning, but both methods let the query optimizer learn by exploration. Neo~\cite{Marcus:NEO:pvldb-2019} is another end-to-end deep neural network-based query optimizer that uses reinforcement learning to guide plan enumeration. Rebopt~\cite{Kaoudi2020MLbasedCQ} is a cross-platform vector-based optimizer which replaces the cost model. In the plan enumeration, it relies on a set of algebraic operations that operate on vectors. $\text{Simpli}^{2}$ is not specific to any execution engine and does not rely on cardinality estimations---or require any training. In fact, it is a simple and generic method whose optimization time is close to zero in any scenario. Despite such extreme simplicity, our evaluation shows similar or better runtime than state-of-the-art methods.

\section{CONCLUSIONS}\label{sec:conclusions}

In this paper, we present the statistics-free $\text{Simpli}^{2}$ algorithm for join ordering. Consequently, $\text{Simpli}^{2}$ does not use cardinality estimates. It is based exclusively on the join graph and type. However, this does not make it a rule-based optimizer. Surprisingly, $\text{Simpli}^{2}$ performs unexpectedly well on the complex JOB benchmark across a variety of databases and configurations. It is at most 16\% slower than the fastest statistics-based method, which changes across configurations. Given its negligible overhead and low maintenance cost, this makes it very appealing for easy adoption in new query optimizers. One may argue that these results are specific to the JOB benchmark and do not hold on larger databases. In that case, we argue that this is a problem with JOB, which fails to properly assess a query optimizer. If a simple algorithm such as $\text{Simpli}^{2}$ performs so well on the benchmark workload, it is evident that JOB is not difficult enough to evaluate challenging join ordering scenarios. Maybe JOB is more suited only for cardinality estimation, not complete query optimization.

\paragraph*{Acknowledgments.}
This work is supported by NSF award number 2008815 and by a U.S. Department of Energy Early Career Award (DOE Career).

\small{
\bibliographystyle{abbrv}

}

\end{document}